\documentstyle[seceq,letter,epsf]{ptptex}

\title{Possible measurement of Quintessence and 
density parameter using strong gravitational lensing events}

\author{Toshifumi {\sc Futamase} and Shijun {\sc Yoshida}}

\inst{Astronomical Institute, Tohoku University, Sendai
980-8578, Japan}

\recdate{
}

\abst{
We propose a possible measurement of the time variability of the vacuum 
energy using strong gravitational lensing events. As an example we 
take an Einstein cross lens HST 14176+5226 and demonstrate that 
the measurement of the velocity dispersion with the accuracy of $\pm$ 
5 km/sec will have a chance to determine the time dependence of the vacuum 
energy as  well as the density parameter with the accuracy of order 0.1 
if one fixes the lens model.
}

\begin{document}

\maketitle

It is now fashionable to assume a non-vanishing value for the
cosmological constant to explain 
some of the observations, such as the magnitude-redshift relation of high-z 
Type Ia supernovae.~\cite{Pe97,Pe99,Ga98,Ri98,Sc98} 
Its origin is not fully understood, but it is associated with the energy 
of the vacuum, and thus called the vacuum energy. 
It is sometimes argued that 
it comes from the energy of the zero-point fluctuation of quantum fields. 
However if it is so, the value expected by a naive theoretical argument
based on the dimensional analysis is about $10^{122}$ larger than 
the observationally suggested value. 
It is also argued that the cosmological constant is identified with the
potential energy of a scalar field. Then the natural energy scale 
will be again much higher than the observationally suggested value 
and one needs to explain such a small energy. Furthermore 
one is then naturally expect that the vacuum energy changes as 
Universe evolves.~\cite{Ca98} 
Thus to reveal the detailed nature of the vacuum energy, i.e, its
existence and time variability, is very important not only for
cosmology but also for fundamental physics.

Several methods have been proposed so far to determine the vacuum
energy, e.g.\ the multiple imaged quasar
statistics~\cite{Fu92,Ma93,Ko96,Ch97,Ch99} and the magnitude-redshift
relation of distant type Ia
supernovae,~\cite{Pe97,Pe99,Ga98,Ri98,Sc98} but very few is 
able to measure its variability.~\cite{Ch00}  
Here we propose a possible method using strong gravitational lensing.
As a matter of fact one of the authors has pointed out that 
Einstein ring system with suitable redshift combinations of the
lens and source can be used as a powerful tool to measure the vacuum
energy.~\cite{FH99} We here point out that the system and other similar 
systems are also powerful to measure the variability as well.
 In the below we restrict ourself to the totally flat cosmology which means 
that  $\Omega_m+\Omega_\lambda=1$, where  $\Omega_m$ and $\Omega_\lambda$ are
the density parameter of matter and normalized cosmological constant at present
time, respectively. This is supported by recent measurements of cosmic
background anisotropy.~\cite{boom,La00,Maxima,Ba00} We also take 
an ``Einstein cross'' gravitational lens system HST14176+5226 as our 
example.

HST 14176+5226 has been discovered with HST as a candidate of 
gravitational lens 
system in 1995,~\cite{Rat95} and subsequent spectroscopic observations 
have provided confirmation that the system is indeed a lens.~\cite{Cra96} 
The elliptical lensing galaxy has a redshift of $z=0.803$ with apparent
magnitude 21.7 in V band and 19.8 in I band. The lensed
source at redshift $z=3.4$ appears to be QSO with an apparent magnitude 
26.2 in V band and 25.7 in I band. The lens model based on a singular 
isothermal
distribution gives a very good fit to the observed images, with
normalized $\chi^2$ of unity.~\cite{Rat99}

The lens model gives us the (tangential) critical line 
which can be written in terms of the  velocity dispersion of the lensing
galaxy and the distance combination $D_{ls}/D_s$,
where $D_{ls}$ is the angular diameter distance from a lens to a source and
$D_s$ is one from the observer to source. 
On the other hand, the distance combination has a strong dependence on 
the vacuum energy $\Omega_\lambda$, but has a week dependence on 
the matter contribution $\Omega_m$ as pointed out by Ref.~\cite{Fu90}.  
Thus if the velocity dispersion of the lensing galaxy as
well as the redshifts of the galaxy and the source are observed, 
the vacuum energy can be in principle well constrained 
when the redshift combinations are appropriate. 
In fact the idea using the tangential critical line as a method to 
determine the vacuum energy has been applied by
Ref.~\cite{Fo97} by observing the decrease of the number density
of  background galaxies in the cluster Cl0024+1654 at the critical
line, and it is argued that the lower bound on the cosmological
constant is obtained assuming the spherical symmetry of the cluster. 
However, the mass distribution of the cluster is likely to deviate from 
spherical symmetry and the effect does depend on the luminosity function 
and the evolution of the background galaxies, which makes this method 
difficult to withdraw any definite conclusion. 
On the other hand, systems like HST14176+5226 have a almost perfect 
symmetry which allows us to have a very good model fitting.

In this Letter we investigate the possibility to 
measure the variability of the vacuum energy using HST 14176+5226.
Although we restrict ourselves to this system, the method can be applied to  
similar systems with symmetrical configurations and appropriate 
redshift combinations, such as Einstein ring system 0047-2808 and 
a quadruple lens system MG0414+0534. We shall describe the vacuum energy 
 as a perfect fluid with the equation of state $p=\omega \rho$ 
with $ -1 \leq \omega \leq 0$. The case $\omega = -1$ corresponds to the
so called cosmological constant. Then the angular diameter distance in 
the totally flat universe with the vacuum energy is given by
\begin{eqnarray}
\label{dist}
&& D(z_1, z_2) = {c\over H_0} {1\over {1+z_2}}  \nonumber\\
 && \times \int^{z_2}_{z_1} 
{{dx}\over {\sqrt{\Omega_m ( 1+x)^3 + ( 1- \Omega_m)(1+x)^{3(1+\omega)}}}}
\, .
\end{eqnarray}

The angular diameter distance comes in 
the lens equation through the 
lens potential which is modeled by an isothermal
ellipsoid model,~\cite{KSB94} 
\begin{equation}
\label{sie}
\Phi = 4\pi \left({\sigma_v\over c}\right)^2 {D_{ls}\over D_s} 
\sqrt{ ( 1- \epsilon) \theta^2_1 + ( 1 + \epsilon) \theta^2_2}
\, ,
\end{equation}
where $\sigma_v$ is the one-dimensional velocity dispersion. 
The ratio $e$ between the minor- and major-axis is related 
with the ellipticity $\epsilon$ as 
\begin{equation}
\label{ellip}
e = \sqrt{{{1 - \epsilon}\over {1 + \epsilon}}}
\, .
\end{equation}
Model fitting  
gives $e = 0.4$ and $\theta_E= 1".489$,~\cite{Rat99} 
where 
\begin{equation}
\label{angle}
\theta_E= 4\pi \left({\sigma_v\over c}\right)^2 {D_{ls}\over D_s} 
\, .
\end{equation}

Knowing the lens and source redshift $z_l$, $z_s$ and $\theta_E$, 
we plot the velocity dispersion $\sigma_v$ in the $\omega-\Omega_m$ plane on 
Figure 1. The region between the dotted vertical lines indicates 
the allowed range of the density 
parameter $0.2 \leq \Omega_m \leq 0.4$ which is suggested from various 
observations concerning clusters.~\cite{Ba99}

\begin{figure}
\begin{center}
    \leavevmode\epsfxsize=7cm \epsfbox{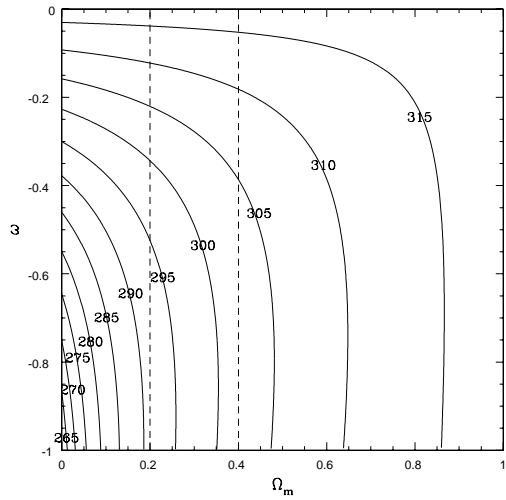}
\end{center}
\caption{ Contours of constant velocity dispersion $\sigma_v$ in 
         the $\omega$--$\Omega_m$ plane. 
         Constant $\sigma_v$ lines are drawn in steps of 5 km/sec for 
         $265 {\rm km/sec} \le \sigma_v \le 315 {\rm km/sec}$. Dotted 
         vertical lines are $\Omega_m=0.2$ and $\Omega_m=0.4$. }
\end{figure}

The figure shows that the velocity diversion 
is highly restricted in the case of $\omega=-1$ (the cosmological constant 
case), namely  $291 \leq \sigma_v \leq 302$ km/sec. 
Thus it shows that the system can be used as a good indicator 
for the existence of the cosmological constant when the observation 
of the velocity dispersion is performed within the accuracy of, 
say, 5km/sec which is achievable by any of 8-10m telescopes 
with reasonable observational time. 
Put it other way, the accurate measurement of the 
velocity dispersion of the system, say $\pm 5$ km/sec  gives us 
a determination of the density parameter $\Omega_m$  with the accuracy of 
order of $\pm 0.1$.

The figure also shows the possibility of measurement of the parameter 
$\omega$, namely the variability of the vacuum energy. If the measured 
value of the velocity dispersion is relatively large 
as 310 $\pm$ 5 km/sec, 
then the $\omega=-1$ solution is inconsistent 
with the expected range of $\Omega_m$. 
Thus it is extremely important to have an accurate measurement 
of the velocity dispersion and modeling of the lensing 
galaxy. Concerning the accurate measurement, recent observation 
by Keck achieved an accurate measurement of the velocity dispersion for
high redshift galaxies.\cite{Dokkum98} 
On the other hand, the modeling of lensing
elliptical galaxies is not easy task because of many theoretical 
ambiguities such as the choice of the dark halo potential 
and the existence of anisotropic velocity dispersions. 
However there have been steady progresses in this direction also. 
Although the isothermal distribution of dark matter halo is not chosen as
the result of $\chi^2$ fitting by varying parameters in the dark matter 
potential in the original modeling of HST 14176+5226\cite{Rat95,Rat99}, 
it has been shown that isothermal distribution is consistent 
with both dynamical consideration\cite{Kor95} and lensing data.\cite{Ko93,Ko96}
Furthermore, dynamical observations and models of elliptical galaxies in 
the local Universe has been studied by Rix et al,~\cite{Rix} 
and dynamical models of local early-type galaxies in SIS halos 
by Kochanek demonstrated that the observed stellar velocity dispersion 
($\sigma_{ob}$) and the velocity dispersion associated with dark matter 
($\sigma_{v}$) have nearly identical values 
using the observed distribution of image separation.\cite{Ko96,Falco99} 
If the dark matter halo has a finite core (within which the density is
roughly constant), $\sigma_{v}$ is increased for a given observed
$\sigma_{ob}$ because the central potential well is shallower. 
There have been some studies on this point and linear dependence on the
core radius $r_c$ is obtained;
\begin{equation}
\sigma_v = \sigma_{ob} \left( A + B {r_c\over r_E}\right)
\end{equation}
where A and B are independent of core radius, and $r_E$ is the Einstein
radius. Kochanek used $A=1$ and $B=2$.\cite{Ko96} But this relation
should be studied more carefully. 
Non-singular lenses tend to produce more highly magnified images
and to produce characteristic images such as radial
arcs so that detailed imaging by large telescopes is definitely 
to have more realistic model of the lensing system.     
It is hoped that further observational and theoretical
progresses will resolve the above ambiguities in near future.

\section*{Acknowledgments}
We thank Takashi Hamana, Masashi Chiba  and Masahiro Takada for valuable 
and critical comments, which
considerably improved this manuscript. We also thank Kazuhiro Yamamoto 
for useful discussion.
S.Y. acknowledges a support from Research Fellowship of the Japan Society 
for the Promotion of Science.

\end{document}